# The highly selective cyclooxygenase-2 inhibitor DFU is neuroprotective when given several hours after transient cerebral ischemia in gerbils

Authors: Eduardo Candelario-Jalil *, Dalia Alvarez, Juana M. Castañeda, Said M. Al-Dalain, Gregorio Martínez-Sánchez, Nelson Merino, Olga Sonia León

Affiliation: Department of Pharmacology, University of Havana (CIEB-IFAL), Apartado Postal 6079, Havana City 10600, Cuba.

*Author to whom all correspondence should be addressed:

**Eduardo Candelario-Jalil, M.Sc.**
**Department of Pharmacology**
**University of Havana (CIEB-IFAL)**
**Apartado Postal 6079**
**Havana City 10600**
**CUBA**
**Tel.: +53-7-219-536**
**Fax: +53-7-336-811**
**E-mail: candelariojalil@yahoo.com**

**Acknowledgements:** The authors are greatly indebted to Dr. Denis Riendeau (Merck Frosst Center for Therapeutic Research, Kirkland, Quebec, Canada) for providing DFU for this study and critically reviewing this manuscript. We thank Dr. Robert Young (Merck Frosst Canada) and Dr. Stefano Govoni (University of Pavia) for their critical comments on the manuscript.

**Abstract**

Several studies suggest that cyclooxygenase-2 contributes to the delayed progression of ischemic brain damage. In this study we examined whether the highly selective cyclooxygenase-2 inhibitor DFU reduces neuronal damage when administered several hours after 5 min of transient forebrain ischemia in gerbils. The extent of ischemic injury was assessed behaviorally by measuring the increases in locomotor activity and by histopathological evaluation of the extent of CA1 hippocampal pyramidal cell injury 7 days after ischemia. DFU treatment (10 mg/kg, p.o.) significantly reduced hippocampal neuronal damage even if the treatment is delayed until 12 h after ischemia. These results suggest that selective cyclooxygenase-2 inhibitors may be a valuable therapeutic strategy for ischemic brain injury.

Brain damage accompanying cardiac arrest and resuscitation is frequent and devastating [24]. The cornu Ammonis 1 (CA1) neurons of the hippocampus are widely regarded as among the most vulnerable in the mammalian CNS to ischemia [11,12,20]. Detailed analysis on the time course of selective loss of pyramidal cells in CA1 hippocampal sector showed that the neuronal death is a slow process taking 2-3 days before presenting the final morphologic outcome [11,20], suggesting that mechanisms that develop after ischemia have an important role in ischemic neuronal demise. The delayed degeneration of neurons provides the opportunity for neuroprotective agents to be administered following the ischemic insult.

There is mounting evidence that post-ischemic cyclooxygenase-2 induction contributes to ischemic brain damage [9,18,19]. It has been known for decades that cyclooxygenase inhibitors decrease ischemic brain injury [17,22]. Recently, it has been reported that cyclooxygenase-2 selective inhibitors prevent both post-ischemic prostaglandin accumulation and ischemic neuronal damage [18,19], suggesting that the beneficial effects observed with non-selective cyclooxygenase inhibitors are probably associated to cyclooxygenase-2 rather than to cyclooxygenase-1 inhibition.

Induction of cyclooxygenase-2 mRNA and protein through activation of AMPA receptors in global ischemia suggests that cyclooxygenase-2 is a mediator of glutamate excitotoxicity [13]. While several genes are induced in the susceptible CA1 neurons after global ischemia, very few reach expression at the protein level [14]. The induced cyclooxygenase-2 mRNA is translated to protein in models of global cerebral ischemia [13,18]. Cyclooxygenase-2 is one of a select few proteins that still remains upregulated in CA1 pyramidal cells even at 3 days after the ischemic insult [13,18], thus preceding the death of these neurons.

In the present study, we examined whether a highly selective cyclooxygenase-2 inhibitor reduces neuronal damage in the gerbil hippocampus when administered several hours after transient global ischemia.

Male Mongolian gerbils (*Meriones unguiculatus*; 60-75 g) were anesthetized with chloral hydrate (300 mg/kg, i.p.) and subjected to both common carotid arteries occlusion for 5 min, exactly as in our previous work [4]. Rectal temperature was monitored and maintained at 37 ± 0.5 °C. In addition, rectal temperature was carefully monitored at 8-h intervals for 3 days of reperfusion. In sham-operated group (n=5), the arteries were freed from connective tissue but were not occluded.

The novel orally active and highly selective cyclooxygenase-2 inhibitor DFU (5,5-dimethyl-3-(3-fluorophenyl)-4-(4-methylsulphonyl) phenyl-2(5H)-furanone) (Merck Frosst, Quebec, Canada) was administered orally (10 mg/kg) 30 min before ischemia and again at 6, 12, 24, 48 and 72 h of reperfusion (n=6). In other groups, DFU was dosed 6 (n=9) and 12 h (n=9) after induction of ischemia, followed by additional single oral doses at 24, 48 and 72 h. This treatment schedule was based on previous reports showing a persistent cyclooxygenase-2 protein expression in CA1 neurons after global ischemia [13,18]. The dose of DFU used in the study was based on a previous work showing maximal antiinflammatory effect [21]. In the ischemic control group (n=7) the animals were treated orally with vehicle (1% methocel) for 3 days.

All animals were tested for spontaneous locomotor activity in an automated activity meter (Model 7401, Ugo Basile, Varese, Italy). Each individual gerbil was monitored for a 30 min period after a 10 min interval for adaptation. Locomotor activity was measured prior to, and 7 days after the ischemic episode. This behavioral test has been previously shown to accurately predict the degree of ischemia-induced hippocampal damage [1].

Following the final motor activity session on the seventh day post-ischemia, the animals were anesthetized with urethane and perfused transcardially with cold saline followed by 4% paraformaldehyde in phosphate-buffered saline (pH 7.4). The brains were removed from the skull and fixed in the same fixative for 24 h. Thereafter, the brains were embedded in paraffin and representative coronal sections (5-µm thick), which included the dorsal hippocampus, were obtained with a rotary microtome. Tissue sections were stained with hematoxylin and eosin. The hippocampal damage was determined by counting the number of intact neurons in the striatum pyramidale within the CA1 subfield at a magnification of 40x [15]. Only neurons with normal visible nuclei were counted. The mean number of CA1 neurons per mm linear length for both hemispheres in three adjacent sections of dorsal hippocampus was calculated for each group of animals. All assessments of histological sections were made by an observer who was unaware of the drug treatment for each gerbil. The data was expressed as a mean value ± SD. Statistical analysis was performed using ANOVA followed by Student-Newman-Keuls post-hoc test and significance refers to results where p<0.05 was obtained.

After 7 days of bilateral carotid occlusion for 5 min, the vehicle-treated ischemic gerbils displayed a characteristic significant increase in locomotor activity with respect to both their own pre-ischemia basal level and the sham group (Table 1). Seven days post-ischemia, all the gerbils which received DFU (10 mg/kg, p.o.) showed a significant reduction (p<0.05) in locomotor activity as compared to vehicle-treated group (Table 1).

Histologic examination of the brain demonstrated marked cell damage in the hippocampal CA1 region in the gerbils treated with vehicle when compared with the sham-operated group. CA1 pyramidal neurons showed pyknosis, eosinophilia, karyorrhexis and chromatin condensation in ischemic gerbils. Figure 1 shows that by 7 days after ischemia, there is a 76% loss of medial CA1 neurons in the vehicle-treated animals (sham = 223 ± 13; vehicle = 55 ± 12 neurons/mm). Administration of the highly cyclooxygenase-2 inhibitor DFU significantly (p<0.05) reduced neuronal damage induced by global ischemia when administered 30 min prior to the 5 min period of carotid occlusion (vehicle = 55 ± 12 *vs* DFU pre-ischemia = 170 ± 16 neurons/mm; Fig 1). We tried to determine whether a cyclooxygenase-2 inhibitor could protect neurons when applied several hours after ischemia. For this purpose we administered DFU 6 and 12 h after the insult. All application schedules led to statistically significant increase in healthy neurons in the CA1 layer (p<0.05) as shown in Fig. 1. Of special interest is the finding that when DFU treatment is delayed until 6 h after ischemia the neuroprotection is comparable to that seen in the group in which DFU treatment began before ischemia (Fig 1). The protective effects of this cyclooxygenase-2 inhibitor are also seen if administrated 12 h after ischemia (Fig. 1). The protection conferred by DFU is not attributable to effects on body temperature because this variable was monitored and did not differ between treated and un-treated groups (data not shown).

The reduction in locomotor activity in DFU-treated groups correlated well with the histopathological evidence of protection against hippocampal damage after ischemia (Table 1 and Fig. 1).

To our knowledge, there is no previous study on the neuroprotective effects of a highly selective cyclooxygenase-2 inhibitor against brain injury induced by global ischemia showing this wide therapeutic window of protection. The present results suggest that cyclooxygenase-2 plays an important role in the deleterious cascade of molecular events which leads to neuronal damage of CA1 pyramidal cells after transient global ischemia.

Cyclooxygenase-2 inhibitors have been previously shown to decrease infarct size in rats after transient middle cerebral artery occlusion [19], to prevent delayed death of CA1 neurons following 30 min of global ischemia in rats [18], to attenuate white matter damage in chronic cerebral ischemia [23] and to reduce ischemic photothrombotic brain injury in rats [6]. The present study confirms the ability of highly selective cyclooxygenase-2 inhibitors to decrease selective vulnerability of CA1 hippocampal neurons and shows that this effect is present even when treatment begins several hours after ischemia. The observation that DFU is protective in ischemic brain injury in gerbils, suggests that the neuroprotective effects of cyclooxygenase-2 inhibitors are maintained across species.

Although the factors responsible for the cytotoxicity of cyclooxygenase-2 in cerebral injury induced by ischemia remain to be clearly defined, it is likely that one of the mechanisms is related to production of reactive oxygen species [16]. In line with this notion, we have recently found that a cyclooxygenase-2 inhibitor is able to reduce hippocampal oxidative damage following excitotoxic brain injury [3], which is a key pathological event of cerebral ischemia. Cyclooxygenase-2 enzymatic activity can also mediate tissue damage by producing pro-inflammatory prostanoids [16]. Interestingly, prostaglandins stimulates calcium-dependent glutamate release in astrocytes [2], thus contributing to excitotoxicity. However, the precise role of prostaglandins in neurotoxicity is controversial because prostaglandin $E_2$ has also been reported to limit the cytotoxic effects of glutamate [5].

On the other hand, pharmacological inhibition of cyclooxygenase-2 has been proven to reduce N-methyl-D-aspartate-mediated neuronal cell death both in vitro [7] and in vivo [9]. Recent investigations have found a potentiation of excitotoxicity in transgenic mice overexpressing neuronal cyclooxygenase-2 [10] and a significant reduction in ischemic brain injury in cyclooxygenase-2-deficient mice [9].

Another mechanism by which cyclooxygenase-2 could contribute to cell death is related to induction of apoptosis. It was found in a previous report that the induction of cyclooxygenase-2 mRNA expression preceded temporally and overlapped anatomically the cellular morphological features of apoptosis in the granule cell layer of the dentate gyrus after hippocampal colchicine injection in rats and cyclooxygenase-2 induction preceded apoptosis in response to serum deprivation in P19 cells [8]. These findings suggest that cyclooxygenase-2 is involved in programmed cell death after brain injury.

Additional experiments are needed to define the relative contribution of the above-mentioned pathogenic events to ischemic cell death related to cyclooxygenase-2 upregulation.

In summary, this study demonstrates that the cyclooxygenase-2 inhibitor DFU has a neuroprotective effect against hippocampal neuronal damage in gerbils with a wide therapeutic window. This may indicate a potential use for cyclooxygenase-2 inhibitors as a valuable therapeutic strategy against cerebral ischemia to target the delayed progression of the damage.

Table 1

Effects of the highly selective cyclooxygenase-2 inhibitor DFU (10 mg/kg, p.o.) on the spontaneous locomotor activity of groups of gerbils. Locomotor activity was measured prior to, and 7 days after 5-min ischemia.

| Groups | Number of movements/30 min/animal | |
|---|---|---|
| | Preischemic values | 7 days after ischemia |
| Sham (n=5) | $54 \pm 12^a$ | $22 \pm 9^a$ |
| Vehicle (n=7) | $52 \pm 8^a$ | $116 \pm 26^b$ |
| DFU pretreatment (n=6) | $60 \pm 10^a$ | $80 \pm 13^c$ |
| DFU after 6 hr (n=9) | $59 \pm 19^a$ | $71 \pm 18^c$ |
| DFU after 12 hr (n=9) | $55 \pm 10^a$ | $79 \pm 17^c$ |

Data are presented as mean ± SD. Values with non-identical superscript are significantly different ($p<0.05$) within the same set. ANOVA followed by Student-Newman-Keuls test.

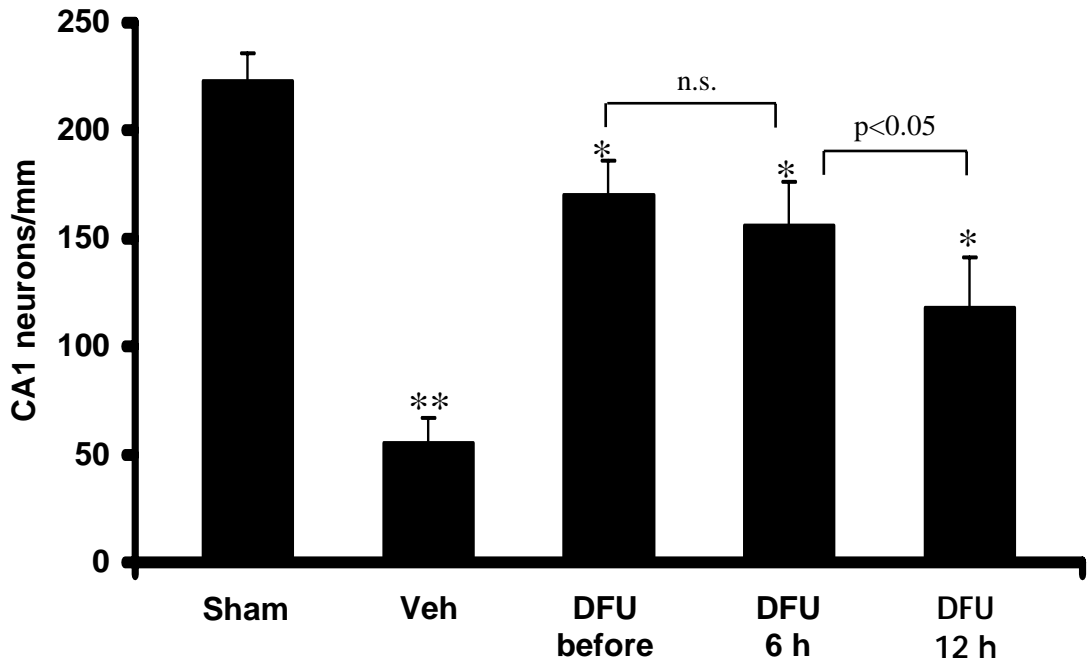

**Fig. 1.** Effect of the selective cyclooxygenase-2 inhibitor DFU (10 mg/kg, p.o.) on the number of surviving cells in the hippocampal CA1 region 7 days after global ischemia in gerbils. Values are mean counts of normal-appearing CA1 neurons ± SD. *$P<0.05$ for the comparison between DFU-treated and vehicle-treated groups. **$P<0.01$ with respect to sham and DFU-treated groups. ANOVA followed by Student-Newman-Keuls post-hoc test. When DFU treatment is delayed until 6 h after ischemia, the neuroprotection is comparable to DFU-pretreated gerbils. Sham, sham-operated (n=5); Veh, vehicle-treated (n=7); DFU before, DFU administration starting 30 min before ischemia (n=6); DFU 6 h, DFU treatment started 6 h after ischemia (n=9); DFU 12 h, DFU administration was delayed until 12 h after ischemia (n=9); n.s., not significant.